\documentclass[a4paper,11pt]{article}
\pdfoutput=1

\usepackage{jcappub}
\usepackage{aas}
\usepackage{orcidlink}
\usepackage{graphicx}
\usepackage{hyperref}
\usepackage{dcolumn}
\usepackage{multirow}
\usepackage{tabularx}
\usepackage{makecell}
\usepackage{booktabs}
\usepackage{xspace}

\usepackage[nameinlink,noabbrev]{cleveref}
\crefname{equation}{Eq.}{Eqs.}
\crefname{section}{Section}{Sections}
\crefname{figure}{Figure}{Figures}
\crefname{table}{Table}{Tables}
\crefname{appendix}{Appendix}{Appendices}
\Crefname{figure}{Figure}{Figures}
\Crefname{equation}{Equation}{Equations}
\Crefname{section}{Section}{Sections}
\Crefname{table}{Table}{Tables}

\usepackage{amsmath}

\usepackage{bm}
\usepackage{tensor}

\begin{document}

\title{Black Holes and Higgs Dark Energy}


\author[1]
{{Steve~P.~Ahlen}\orcidlink{0000-0001-6098-7247},}
\affiliation[1]{Department of Physics, Boston University, 590 Commonwealth Avenue, Boston, MA 02215}

\author[1]
{{James W. Rohlf}\orcidlink{0000-0001-6423-9799},}

\author[2]{{Gregory~Tarl\'e}\orcidlink{0000-0003-1704-0781}}
\affiliation[2]{Department of Physics, University of Michigan, 450 Church St., Ann Arbor, MI, 48109}
\emailAdd{gtarle@umich.edu}

\arxivnumber{...}

\abstract{Black holes, dark energy, and the Higgs field are all currently  established, exciting, and mysterious, each in its own way. Cosmological data show that dark energy may evolve with time. The electroweak phase transition during stellar collapse can provide a mechanism via the Higgs field for dark energy to be trapped inside black holes at the time of their formation. Using the Oppenheimer-Snyder model of collapse, we calculate the total matter and dark energy densities in a black hole, to be in the ratio of 2 to 1 at the start of collapse. The solution for the scale factor {\it a}({\it t}) is a cycloid with a collapse time of 57 \textmu s.  If black holes are cosmologically coupled and grow in mass as the universe expands, they can account for the evolution and quantity of the dark energy of the universe.

}




\maketitle

\section{Introduction}\label{sec1}

\begin{figure}
  \centering
  \includegraphics[width=5. cm]{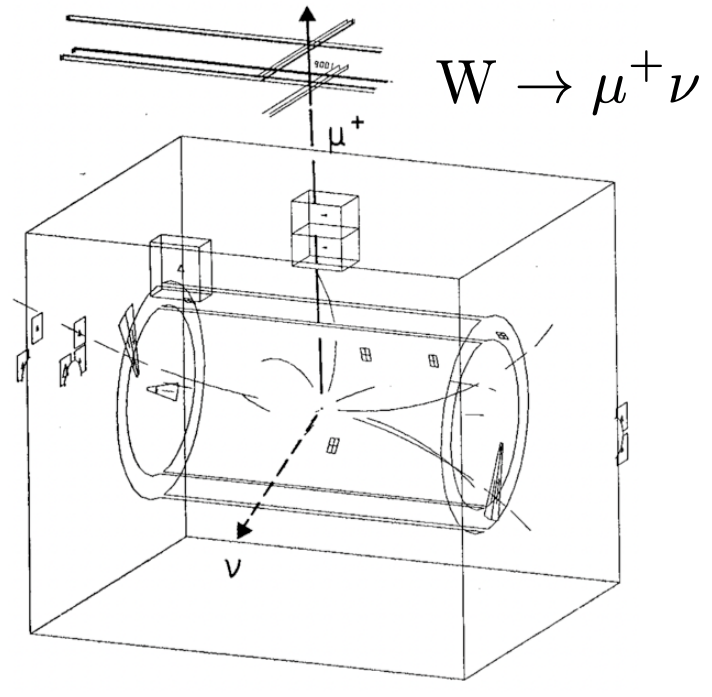}
  \includegraphics[width=7.5 cm]{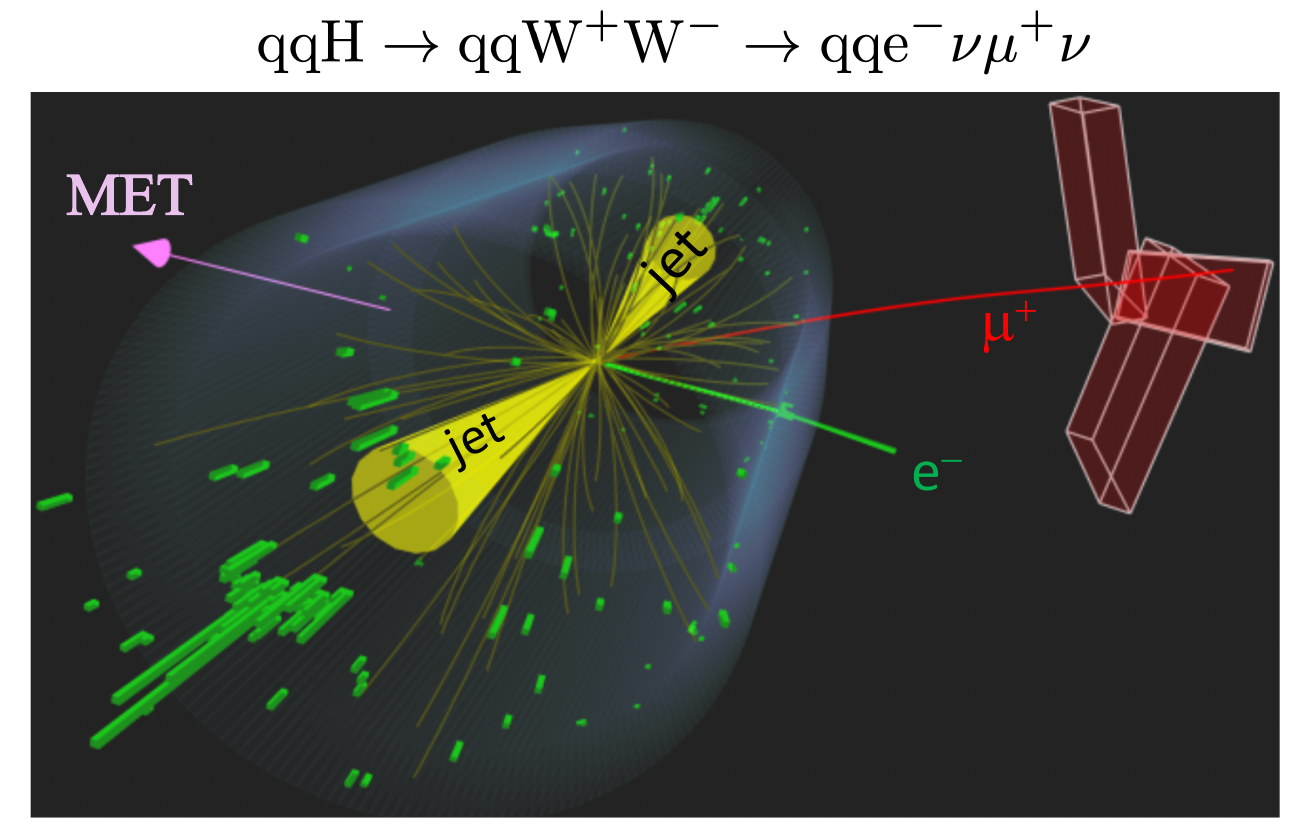}
  \caption{The W boson has been a key ingredient
 in the search and discovery of the Higgs boson and measurement of its properties. (Left) Event from Underground Area 1 at CERN \cite{Rohlf:1984dir}  showing $\rm q{\bar q}\to W\to \mu^+\nu$. The muon is the only high-momentum particle detected. The presence of the $\nu$ is inferred from missing transverse energy \cite{UA1:1983crd}. (Right) Event from the Compact Muon Solenoid showing the decay of the Higgs boson into W pairs \cite{CMS:2022uhn}. The Higgs is produced by vector-boson fusion, the radiation of W or Z (V) from energetic quarks, $\rm qq\to qqVV\to qqH \to qqWW \to qqe^-\nu\mu^+\nu$. The yellow cones show the quarks after fragmentation (jet) and the green and red lines are the charged leptons from $\rm W\to e\nu$ and $\rm W\to \mu\nu$ decays, respectively. The neutrinos are not directly detected, but inferred from missing transverse energy (MET, pink arrow).  These, and measurements that followed, form the basis of our understanding of the Higgs mechanism used to generate DE in collapsing BHs under conditions of density and temperature comparable to those seen in collisions in the terrestrial accelerator environment.
    }\label {fig-HtoWW}
\end{figure}

This paper addresses a possible connection between three of the most exciting, yet mysterious, discoveries in particle physics and cosmology: black holes (BHs), dark energy (DE), and the Higgs field ($\Phi$). In 1939, the BH mystery (for the history of the first use of the term, see \cite{ann_nelson,tom_siegfried}) advanced from a theoretical curiosity \cite{Schwarzschild:1916uq} to a physical plausibility when J. Robert Oppenheimer and his student, Hartland Snyder, developed a model (OS) \cite{PhysRev.56.455} of collapsed stars consistent with Einstein's equations of general relativity \cite{einstein,Einstein:1916vd}.  
Although originally met with widespread skepticism, the main features of OS collapse were confirmed with detailed computer simulations by Colgate and Johnson in 1960 \cite{PhysRevLett.5.235}. A decade later, Cygnus X-1 was identified as the first BH candidate \cite{1972Natur.235...37W}, and the OS model continues to be discussed, used, and extended (see for example, Shojai, Sadeghi and Hassannejad \cite{Shojai_2022}).
The DE mystery (see \cite{1999ASPC..165..431T} for a history of the name) arose from the discovery in 1998 of an
unexpected acceleration of the expansion of the Universe \cite{Riess_1998, Perlmutter_1999} and has since been supported by a variety of observations, including formation of large-scale structure, fluctuations of the cosmic microwave background (CMB) radiation, and baryon acoustic oscillations (for an overview, see \cite{Blinnikov_2019}). 
The Higgs theory was developed in 1964 \cite{Higgs:1964ia,Higgs:1964pj,Englert:1964et,Guralnik:1964eu} 
(see also \cite{Anderson2015})
and the corresponding particle, after a decades-long search, was finally discovered at the CERN Large Hadron Collider (LHC) in 2012 \cite{ATLAS:2012yve,CMS:2012qbp}. 
The vector bosons, W \cite{UA1:1983crd} and Z \cite{UA1:1983mne}, have served as fundamental building blocks in the identification and measurement of the Higgs boson (see Fig. \ref{fig-HtoWW}) \cite{,Rohlf:1984dir,Rohlf:2004nf}, and upgrades to the LHC detectors for operation at high luminosity are centered around maintaining the ability to trigger on the W \cite{CMS:2017lum,2137107}.
Measurement of the properties of the Higgs boson is an ongoing major effort at the LHC \cite{CMS:2022dwd,ATLAS:2022vkf} and a central physics goal in the design of future accelerators \cite{FCC:2018byv,Accettura:2023ked}.

Scalar fields were used to invent inflation \cite{Guth:1980zm,LINDE1983177}, and have long been considered as a leading candidate for the source of DE. Efforts to find other scalar fields beyond $\Phi$ have produced null results \cite{CMS:2024yiy,ATLAS:2022eap}. As the only known scalar field in particle physics, there have been attempts to connect $\Phi$ with inflation \cite{BEZRUKOV2008703} and even interesting variations to connect it with DE (for example, see \cite{Rinaldi:2014yta,Usman:2019gax}).  
Recent data from the Dark Energy Spectroscopic Instrument (DESI) \cite{DESI2024_VI} provide robust measurements of DE.
Taken together, DESI plus CMB data from Planck  \cite{Planck18_VI,PlanckLensing2020} and SNe data \cite{Rubin2023Union, descollaboration2024dark,Brout_2022}  suggest that dark energy may be emerging from matter sometime between cosmic ``dawn" and cosmic ``noon," the period of time when black holes were being formed from massive stars. 
In fact, it has been observed that the evolution of DE seems to follow the formation of stars \cite{Croker:2024jfg}.
This may be interpreted as evidence for DE in stellar collapsed BHs as the cause of the accelerated expansion of the Universe, originally proposed to proceed by means of cosmologically coupled BHs (CCBHs) by Farrah et al. \cite{Farrah_2023}.
With the exception of a few comments in a paper by Mboyne and Kazanas \cite{Mbonye:2005im}, noting that this paper was written several years before the discovery of the Higgs boson, no study that we are aware of has attempted to connect DE from $\Phi$ contained within CCBHs. 
The formation of a BH provides the only environment in the late Universe in which a time-reversed expansion of the early Universe can occur. It is inevitable that $\Phi$ is produced within BHs, making it perhaps the most logical candidate for DE.

In the following, we present a highly simplified model of a compact object made from the gravitational collapse of a stellar remnant, which contains a quark-gluon plasma (QGP) undergoing an electroweak phase transition (EWPT) on its path to becoming a BH. We find that the BH contains a considerable amount of mass due to the Higgs potential energy density. $\Phi$ was invented to explain the W and Z masses, which are zero in the basic theory \cite{griffiths,Zee2010,kane}. $\Phi$ has two components, each of which has a mass and each of which can be written as a complex field. The vacuum is the state for which one of the two components is zero, and the scalar field is studied by examining its behavior relative to the vacuum. To make $\Phi$ invariant under local gauge transformations, one replaces the derivatives in the Lagrangian density with terms involving massless gauge fields $A_{\rm \mu}$, as used in quantum mechanics to replace the momentum operator for electromagnetism. The resulting Lagrangian density then consists of free-field parts for a massive Higgs field, a massless Higgs field, a free gauge field which has acquired a mass term, and 10 terms involving interactions of the fields with each other and with themselves. One of the interaction terms is physically meaningless and can be eliminated by a transformation of the real and imaginary Higgs components. This results in the elimination of the massless Higgs term, leaving a massive Higgs component, a massive gauge field, and four interaction terms.

It is now known that the interaction with $\Phi$ gives mass not only to the W and Z bosons, but also to quarks, electrons, muons, and tau leptons (not neutrinos \cite{Murayama:2006qb}, although they are now known to have mass). For cosmology, we take the energy density ($\rho$) of $\Phi$ in natural units ($\hbar = c = 1)$ to be
\begin{equation}\label{eq:myequation}
\rho=\frac{\dot \Phi^2}{2}+V(\Phi),
\end{equation}
where the first term on the right is the kinetic energy density with dots denoting time derivatives, and the second is the potential energy density \cite{huang}. The pressure ($P$) is
\begin{equation}\label{eq:myequation2}
P=\frac{\dot \Phi^2}{2}-V(\Phi).
\end{equation}

Using conservation of energy, it is easily found that the ``equation of motion" for a spatially uniform $\Phi$ in an expanding or contracting universe is
\begin{equation}\label{eq:myequation3}
{\ddot \Phi}+3H\dot\Phi+\gamma\dot \Phi=-\frac{dV}{d\Phi} ,
\end{equation}
where  $H$ is the Hubble parameter, $\gamma$ is a damping constant due to radiation and interactions with particles, and $a(t)$ is the scale factor, with
\begin{equation}\label{eq:myequation4}
H=\frac{\dot a}{a} .
\end{equation}
The equation of motion is similar to that of a damped harmonic oscillator. One can study problems involving $\Phi$ as if a small object was sliding on a hilly surface in the shape of the Higgs potential. The potential is a function of $\Phi$, and the negative gradient of the potential acts as a horizontal force. The friction of the surface can be caused by the change of the scale factor of the Universe (the Hubble friction), radiation of energy in the form of particle mass, or other dissipative interactions. Since the kinetic energy term of the Higgs energy density is believed to be small because of the various damping mechanisms, the equation of state of the Higgs energy density is $w =P/\rho= -1$, making the Higgs energy density a DE candidate. If the Higgs is the source of DE, it must be locked up in discrete objects of small volume, since, if it is uniformly distributed throughout the Universe, it exceeds the measured DE density by 54 orders of magnitude \cite{annurev:/content/journals/10.1146/annurev.aa.30.090192.002435,RUGH2002663}. 
CCBHs as described by Farrah et al. \cite{Farrah_2023} and Croker et al. \cite{Croker:2024jfg}, may be such Higgs compact objects (HCOs), and provide a mechanism through which DE evolves with time, as suggested by the DESI data, in contradiction with the standard model of cosmology \cite{1999ASPC..165..431T}.

\section{Black-Hole Metrics}\label{sec2}

We will consider the core collapse of a spherically symmetric object with mass ($M$) 
greater than the Tolman-Oppenheimer-Volkoff (TOV) limit, beyond which collapse to a BH is expected to occur \cite{1939PhRv...55..364T,1939PhRv...55..374O,Rezzolla_2018}.
It should be noted, however, that the TOV limit cannot be used to establish a lower-bound BH mass for CCBH or any other non-singular and/or horizonless BH models.
We follow Weinberg's discussion of the OS model \cite{weinberg} which uses two metrics, one interior to the collapsing object and one exterior.  The Schwarzschild metric is used for the external solution:

\begin{equation}\label{eq:myequation5}
ds^2 = -\left(1-\frac{r_{\rm s}}{r}\right)c^2 d{ t}^2+
\left(1-\frac{r_{\rm s}}{r}\right)^{-1} d{ r}^2
+r^2d\Omega^2,
\end{equation} where $r^2d\Omega^2 = r^2(d\theta^2+\sin^2\theta d\varphi^2)$, $s$ is the space-time interval in terms of time $t$, distance $r$, and polar ($\theta$) and azimuthal ($\varphi$) angles, 
and $r_{\rm s}=2MG/c^2$ is the Schwarzschild radius in terms of 
the speed of light $c$, and the gravitation constant $G$.
It should be understood that the Schwarzschild metric does not match the Robertson-Walker (RW) boundary conditions at spatial and temporal infinity, and thus is {\it not} a solution to Einstein's equations in our expanding universe.  Since we are dealing with collapse times much smaller than the Hubble time, the Schwarzschild solution will be an excellent approximation to the true metric, although we note that on cosmological time scales, it will not exhibit the cosmological coupling required for HCOs to be the DE.

A RW metric is also used for the internal solution:

\begin{equation}\label{eq:myequation6}
ds^2=-c^2dt^2+a(t)^2\left[\frac{dr^2}{1-(r/R_0)^2}+r^2d\Omega^2 \right],
\end{equation}
where $a(t)$ is the time-dependent scale factor. Unlike the RW metric for the Universe, the RW metric for the internal BH solution has positive spatial curvature, with $R_0$ being the radius of curvature.

\section{Electroweak Phase Transition}

The internal solution is found using the Friedmann equation

\begin{equation}\label{eq:myequation7}
\frac{\dot a^2}{a^2}=\frac{8\pi G}{3c^2}\rho(a)-\frac{c^2}{a^2R_0^2},
\end{equation}
where $\rho(a)$ is the scale-dependent total energy density, representing the sum of that due to matter, radiation, and $\Phi$.

\begin{figure}
  \centering
  \includegraphics[width=10 cm]{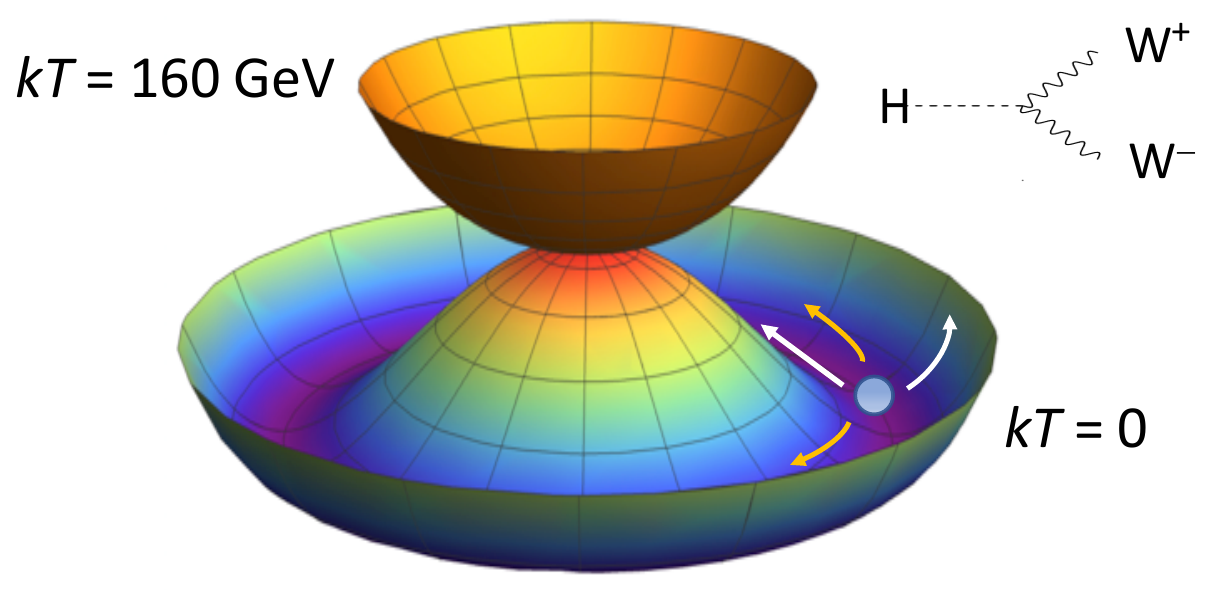}
  \caption{Higgs potential for a complex scalar field with $kT=0$ (rainbow) and $kT=T_{\rm c}$ at the onset of the EWPT (orange). The coupling of the Higgs boson to the W boson during the collapse of a compact object, $\rm H\to WW$, provides a mechanism for dark energy to be captured by a BH.  The location of the ball represents a quantum phase angle at the minimum of the potential chosen during spontaneous symmetry breaking. Motion along the azimuthal direction represents a yet-to-be observed massless Goldstone boson mode \cite{Zee2010}. Longitudinal motion represents energetic, radial excitations of $\Phi$. This is the part that has been measured with particle physics experiments at accelerators. The distance of the ball from the axis of symmetry of the ``champagne bottle bottom" and the ``cup" is the magnitude of $\Phi$. The motion of the ball on the projected plane when viewed from above, represents variations of $\Phi$. The ball moves in accord with the equation of motion of $\Phi$, which is very similar to the equation of motion for an object moving on a hilly surface in a gravitational field. At $T = 0$, various damping mechanisms cause the ball to settle in the position where the potential is minimum. The value of $\Phi$ at this location is referred to as the vacuum expectation value. The non-zero value of $\Phi$ causes it to interact with particles, giving them mass terms in the Lagrangian. At temperatures above 160 GeV, $\Phi$ is zero and particles therefore do not have mass. The energy of their mass has been transferred to the Higgs potential. For large temperatures, the shape of the potential is like that of the cup, and the ball representing $\Phi$ is sitting at the bottom of the cup, with small amplitude oscillations due to interactions with massless particles.
    }\label {fig-hat}
\end{figure}

For the dependence of the Higgs energy on scale factor, which is usually expressed in terms of temperature $T$, we assume that $T$ is proportional to $1/a$, as seen in black-body radiation. The temperature-dependent expression for the Higgs potential ($V$) is

\begin{equation}\label{eq:myequation8}
V=\lambda[(\Phi^2-\nu^2/2)^2+\nu^2\Phi^2(T/T_c)^2] ,
\end{equation}
where $\lambda$ in terms of the Fermi constant ($G_{\rm F}$) and the Higgs mass ($m_{\rm H})$ is $G_{\rm F}m_{\rm H}^2/\sqrt{2}\approx 0.13$, and the minima of the potential at $T=0$ is $\pm \nu/\sqrt{2}$, with $\nu = \sqrt{1/(\sqrt{2} G_{\rm F})}$ = 246 GeV.
The critical temperature for the EWPT is $kT_{\rm c} \approx 160$ GeV. The potential (Fig. \ref{fig-hat}) is particularly clearly explained by
Melo \cite{Melo_2017} and also Huang \cite{huang} who interpreted the symmetry in terms of a quantum phase. We write $(T/T_{\rm c})^2 = (a_{\rm c}/a)^2$, where $a_{\rm c}$ is the scale factor that corresponds to the critical temperature.

\begin{figure}
  \centering
  \includegraphics[width=12 cm]{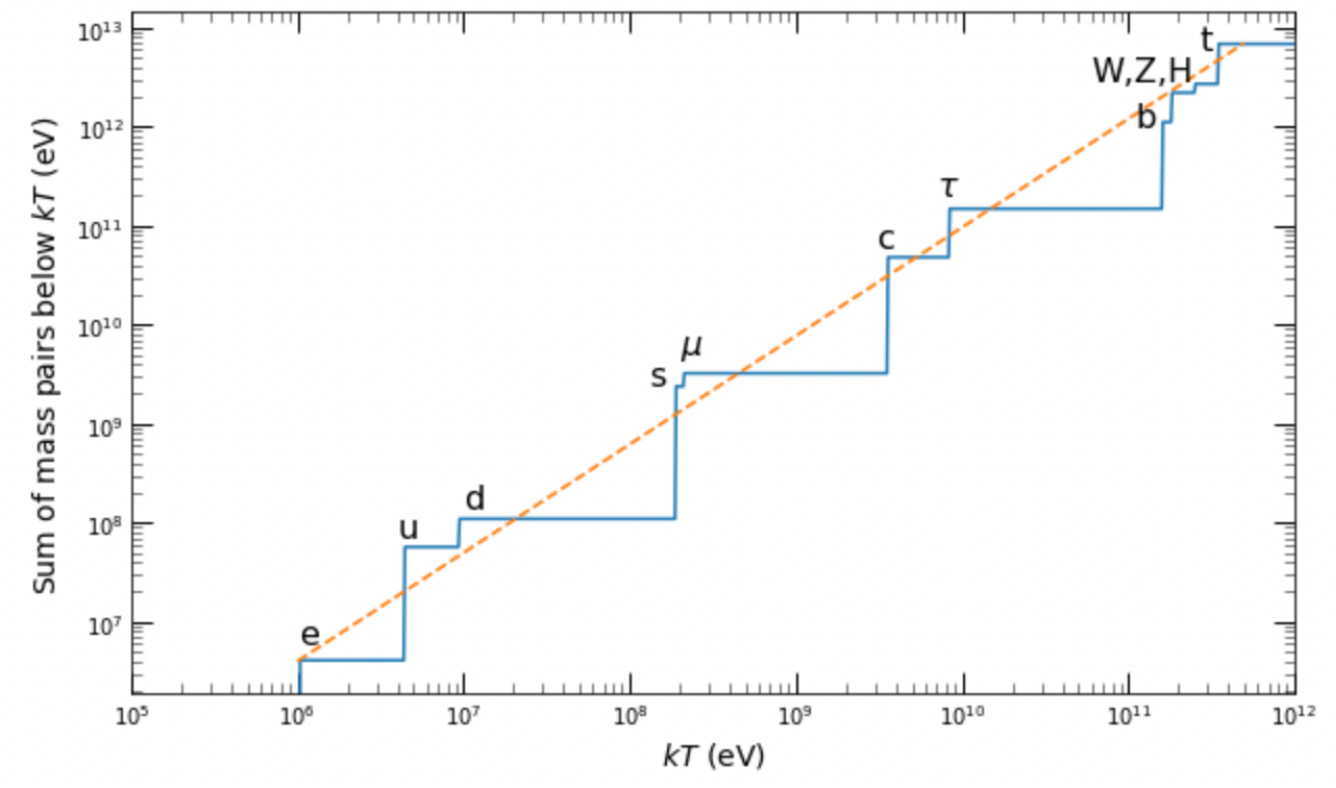}
  \caption{Cumulative sum of particle-pair masses that become above threshold as a function of $kT$, taking spin and color into account. The rich particle spectrum gives a growth curve, approximated by the dashed line, with few flat regions.
    }\label {fig-spectra}
\end{figure}

It is evident that there is an overestimation of $V$ at low temperature because it does not account for the activation of particle species as the QGP is approached. Chaudhuri and Dolgov \cite {chaudhuri_2018} explain the impact of this on $V$ in some detail, but basically one finds that the contribution to $V$ for particle mass pairs below threshold is exponentially suppressed. Figure \ref{fig-spectra} shows the activated particle species as a function of temperature that is remarkably smooth (with few flat regions) over many orders of magnitude. Taking advantage of this feature, we insert another power of $T/T_{\rm c}$ to obtain an approximate expression for $V$,

\begin{equation}\label{eq:myequation9}
V=\lambda[(\Phi^2-\nu^2/2)^2+\nu^2\Phi^2(T/T_c)^3].
\end{equation}

We assume that the collapse begins at the Schwarzschild radius $r_{\rm s} = 2MG/c^2 $ (about 9 km for $M= 3 M_\odot$) with $a = 1$. This seems reasonable since, as shown by Weinberg \cite{weinberg}, a spherically symmetric star can exist in a stable static state only if its radius is greater than $9r_{\rm s}/8$. Accordingly, we use 

\begin{equation}\label{eq:myequation10}
kT|_{a=1} = kT_{\rm c}a_{\rm c} = (160 ~\rm GeV)a_{\rm c}.
\end{equation}
As described below, this corresponds to $a_{\rm c} \approx 2.3 \times 10^{-4}$ and $kT\approx 37$ MeV at $a=1$ for $M= 3 M_\odot$.

\section{Methods}
For $1 > a > 0.1$, we assume that baryons dominate over radiation and that they are governed by a soft equation of state for which the pressure is small compared to the energy density. For the initial stages of the collapse, which we assume begins with nuclear matter, we get guidance from numerous detailed studies of neutron stars. In particular, as discussed by Lattimer \cite{10.1063/1.4909560}, the birth temperature of neutron stars is as large as 
$kT \approx 60 ~\rm MeV$,
and Sumiyoshi, Kojo, and Furusawa \cite{Sumiyoshi} show that the transition between a neutron star and BH is at a temperature close to 30 MeV. We choose the initial temperature of the collapse to maximize the Higgs potential energy of the HCO without violating the second Friedmann equation (the so-called acceleration equation), 

\begin{equation}\label{eq:myequation11}
\frac{\ddot a}{a}=-\frac{4\pi G}{3c^2}(\rho+3P)=-\frac{4\pi G}{3c^2}(\rho_{\rm m}-2\rho_{\rm H}) ,
\end{equation}
where  $\rho_{\rm m}$ and $\rho_{\rm H}$ are the matter and Higgs energy densities, respectively.
This seems to require that the matter density be at least twice the DE density to avoid having a positive acceleration of the HCO at the initial time of collapse ($t = 0$). However, this limit may not actually apply since it requires that the assumed scaling behavior of $a(t)$ be rigorously correct and that the flow of heat between the QGP and $\Phi$ is instantaneously and exactly balanced at all times.

Since we neglect pressure, the baryon energy density scales as $1/a^3$, which is proportional to the number density of particles. This is clearly an approximation since 37 MeV is a high temperature. The electron-positron-photon plasma will consist of relativistic particles, which would have $P=\rho/3$, and would lead to a dependence on scale factor of $1/a^4$. Oppenheimer and Snyder \cite{PhysRev.56.455} justified this approximation with the statements ``We have been unable to integrate these equations except when we place the pressure equal to zero... with $P=0$ we have the free gravitational collapse of matter. We believe that the general features of the solution obtained in this way give a valid indication even in the case that pressure is not zero." Weinberg's justification \cite{weinberg} was ``a proper treatment of gravitational collapse would be prohibitively complicated for this book. In order to get some feeling for what can happen during collapse, we consider only the simplest case, the spherically symmetric collapse of ``dust" with negligible pressure."

We are working under the same constraints as Oppenheimer and Snyder, and Weinberg. However, it should be added that the approximation is not as extreme as it might seem. Tawfik and Mishustin  have recently calculated the equation of state for QCD and electroweak phase transitions \cite{nasser2019equation}. They find that $\rho /T^4$ is constant for kT from 1 to 10 GeV and above 100 GeV, suggesting $1/a^4$ dependence for these regions. For $kT < 250$ MeV, they find that P is small compared to $\rho$, suggesting $1/a^{3.48}$ dependence. This corresponds to a soft equation of state for nuclear matter, as might be the case for certain types of neutron stars. Finally, they found that from 10 GeV to 100 GeV, $\rho$ increases more rapidly than P. The physical basis for this is probably associated with the approach to the EWPT from low temperature. Once the collapse of a compact object reaches $10^{-2} > a(t)>  10^{-4}$ there is a copious production of W, Z, H, and top quarks,
which are instrumental in driving the EWPT in the high-temperature direction \cite{PhysRevLett.25.895}. We can think of this process as the conversion of the rest mass of the particles into $V$. These heavy particles will be mostly non-relativistic and, like low-energy baryons in the neutron star, can also be approximated as ``dust" with energy density going as $1/a^3$. The region $0.1 > a > 0.01$ will be dominated by the QGP equation of state, which scales approximately as $1/a^4$. For this region, we assume that the results will not be significantly degraded if we use $1/a^3$.

Using
$\rho_{\rm m}/c^2=c_0/a^3 $
and 
$\rho_{\rm H}/c^2=\rho_0/a^3 ,$
the Friedmann equation becomes

\begin{equation}\label{eq:myequation12}
\frac{\dot a^2}{a^2}=\frac{8\pi G}{3}\frac{(c_0+\rho_0)}{a^3}-\frac{c^2}{a^2R_0^2} .
\end{equation}
Initially, the collapse will reach a neutron star configuration with a radius of about 10 km, where the nuclear forces slow the collapse. After a short time, gravity overwhelms the nuclear forces and the collapse resumes. We can then establish the initial conditions at the Schwarzschild radius to be
${\dot a}(0)=0$
and
$a(0)=1$.
Then

\begin{equation}\label{eq:myequation13}
\frac{8\pi G(c_0+\rho_0)}{3}=\frac{c^2}{R_0^2}
\end{equation}
and

\begin{equation}\label{eq:myequation14}
{\dot a}^2=\frac{8\pi G(c_0+\rho_0)}{3}\left(\frac{1}{a}-1\right) .
\end{equation}
Defining $\tau = t \sqrt{8\pi G(c_0+\rho_0)/3}$,

\begin{equation}\label{eq:myequation15}
\left(\frac{da}{d\tau}\right)^2={\dot a}^2\left(\frac{3}{8\pi G(c_0+\rho_0)}\right)=\frac{1}{a}-1.
\end{equation}
The solution is a cycloid with

\begin{equation}\label{eq:myequation16}
a=\frac{1}{2}(1+\cos\psi), 
\end{equation}

\begin{equation}\label{eq:myequation17}
\tau=\frac{1}{2}(\psi+\sin\psi) ,
\end{equation}
and

\begin{equation}\label{eq:myequation18}
\tau = t \sqrt{\frac{8\pi G(c_0+\rho_0)}{3}} .
\end{equation}

Weinberg \cite{weinberg} matches the inside and outside solutions and finds that $R_0$ is equal to the Schwarzschild radius. Then

\begin{equation}\label{eq:myequation19}
M=\frac{4\pi r_{\rm s}^3}{3}(c_0+\rho_0).
\end{equation}

\section{Discussion}
\begin{figure}
  \centering
  \includegraphics[width=12 cm]{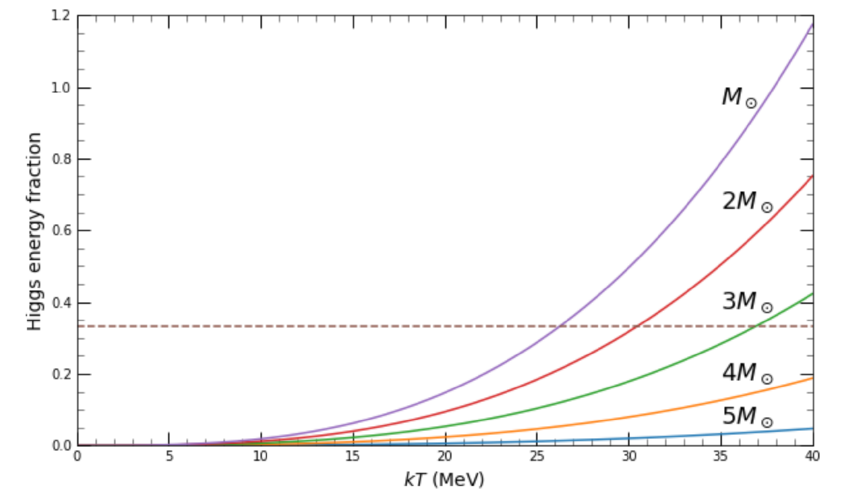}
  \caption{Fraction of energy in the Higgs field vs. $kT$ of the compact object for varying multiples of the solar mass. The dashed line indicates the threshold below which the Higgs fraction must exist to capture Higgs energy in the resulting black hole as calculated from the Friedman acceleration equation.
    }\label {fig-energy}
\end{figure}

Figure \ref{fig-energy} shows the Higgs energy fraction as a function of $kT$ for several values of $M$, with the dashed line indicating the 2 to 1 ratio of matter to Higgs energy density that was discussed above.  If we assume that the Higgs field is equal to its vacuum expectation value, the above results give

\begin{equation}\label{eq:myequation20}
\rho_0 \approx 0.003 \rm ~GeV^4
\end{equation}
and

\begin{equation}\label{eq:myequation21}
c_0 \approx 0.006 \rm ~GeV^4
\end{equation}
in natural units, with
dark energy comprising about 1/3 of the mass of the HCO. Using the numbers from Croker et al. \cite{Croker:2024jfg}, one finds that, with an order of magnitude more mass due to accretion and/or cosmological coupling (which leads to the growth of the mass of the HCOs as the Universe expands), one can account for the dark energy. Of course, this requires that the Universe outside the HCOs can sense and do a spatial average of the equation of state for the internal composition of the HCOs, just as it does for the mass, so that it accelerates in response to this influence.

\begin{figure}
  \centering
  \includegraphics[width=12 cm]{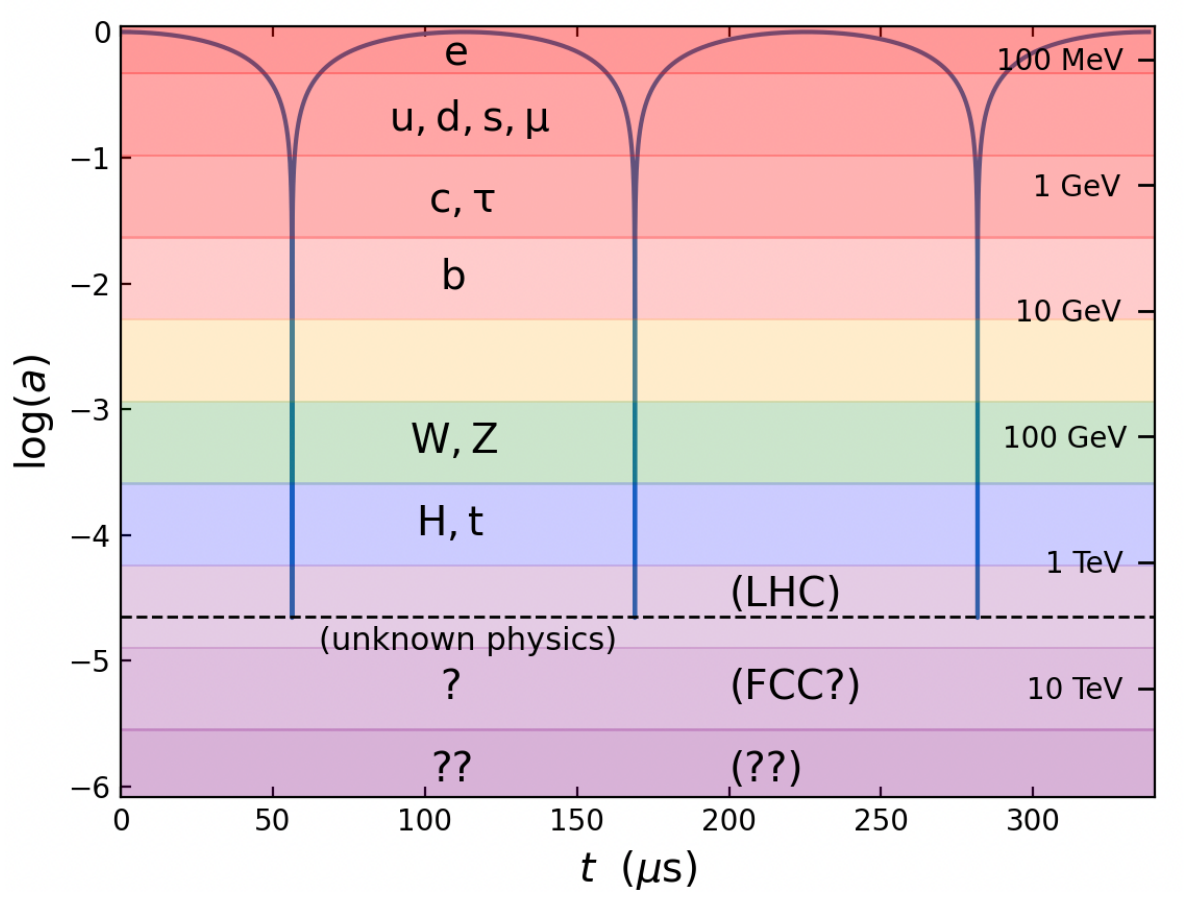}
  \caption{The solution to the Friedman equation with $a^{-3}$ dependence of the matter density and  Higgs energy density is a cycloid shown as the scale factor $a$ vs. time $t$. The HCO collapses starting at $kT=60\rm ~MeV$ from $r_{\rm s} = 9$ km to 0 in 57 $\mu$s. The thresholds for particle production are indicated for values of $a$ along with the $kT$ scale on the right.  Particle physics experiments have explored the region down to $a\approx 2\times 10^{-5}$.  There are two notable energy gaps in the particle spectrum, corresponding to the large energy increase needed to reach the W and Z (yellow region) and the lack of any massive particles observed so far on the TeV scale in 15 years of ongoing LHC operation (top purple region). The question marks indicate the regions to be explored by future accelerators and/or astrophysical measurements.
    }\label {fig-cycloid}
\end{figure}

Figure \ref{fig-cycloid} shows the cycloid solution, over several cycles, using the above parameters. It is not clear whether the actual situation involves only the first half-cycle of gravitational collapse to a singularity, or whether there are periodic bounces \cite{huang}.
The actual behavior depends on the unknown nature of the approach to the singularity.

The question of whether the internal behavior of the HCO is cyclic also relates to the question of entropy. It seems to be the consensus that the EWPT is a crossover (second-order) phase transition and that a more violent first-order transition is ruled out unless new physics is discovered beyond the standard model. Since a crossover transition is generally characterized as being reversible, one would think that this would suggest that entropy is conserved through the transition. Reversibility implies  
that the entropy density ($s$) is given by

\begin{equation}\label{eq:myequation22}
s=\frac{\rho + P}{T}
\end{equation}
and that the co-moving entropy density, $(\rho + P)a^3/T$, is conserved. This is easily seen to be the case for cold non-relativistic dust ($sa^3$ is proportional to a constant divided by a low, constant temperature), radiation ($sa^3$ is proportional to $1/(aT)$) or the Higgs potential ($\rho + P=0$). However, if the various sources of energy density are mixed and interact, things are more complicated. For example, if the energy density of the Higgs potential is converted entirely into radiation, the entropy increases, and vice versa. As expected, entropy would increase during expansion, and decrease during contraction \cite {chaudhuri_2018}. Similar behavior is expected to occur within a cyclic HCO. Only if the cyclic HCO terminates on an expanding part of the cycle, and the BH exploded similar to the launch of our universe, would entropy be conserved and violation of the second law of thermodynamics be averted.

\section{Acknowledgement}
JR acknowledges support from the US Department of Energy under grant DE-SC0016021.

\begin{appendix}
\end{appendix}

\bibliography{main}{}
\bibliographystyle{JHEP}

\end{document}